\documentclass[aps,prl,superscriptaddress,twocolumn,floatfix]{revtex4}

\usepackage{graphicx,latexsym,color}

\begin{document}


\title{Grain growth beyond the Mullins model, capturing the complex physics
behind universal grain size distributions}

\author{R. Backofen}
\affiliation{Institut f\"ur Wissenschaftliches Rechnen, Technische Universit\"at Dresden, 01062 Dresden, Germany}
\author{K. Barmak}
\affiliation{Department of Applied Physics and Applied Mathematics, Columbia University, New York, NY 10027, US}
\author{K.R. Elder}
\affiliation{Department of Physics, Oakland University, Rochester, MI 48309-4487, US}
\author{A. Voigt}
\affiliation{Institut f\"ur Wissenschaftliches Rechnen, Technische Universit\"at Dresden, 01062 Dresden, Germany}


\date{\today}

\begin{abstract}
Grain growth experiments on thin metallic films have shown the geometric and
topological characteristics of the grain structure to be universal and
independent 
of many experimental conditions.  The universal size distribution, however, is
found to differ both qualitatively and quantitatively from the standard Mullins
curvature driven model of grain growth; with the experiments exhibiting an
excess of small grains (termed an "ear") and an excess of very large grains
(termed a "tail") compared with the model.
While a plethora of extensions of the Mullins model have been proposed to
explain these characteristics, none have been successful. In this work, 
large scale simulations of a model that resolves the atomic scale on diffusive
time scales, the phase field crystal model, is used to examine the complex 
phenomena of grain growth. The results are in remarkable agreement with the
experimental results, recovering the characteristic "ear" and "tail"  features
of the experimental grain size distribution. The simulations also indicate that
while the geometric and topological characteristics are universal, the dynamic
growth exponent is not. 
\end{abstract}

\maketitle


Most metals, ceramics and minerals are polycrystalline materials containing
grains of different crystal orientation. The size, shapes and arrangments of
these grains strongly affect macroscale material properties, such as fracture,
yield stress, coercivity and conductivity.  In magnetic systems, for example,
the coercivity (or magnetic 'hardness') can change by four or five orders of 
magnitude with a change in grain size \cite{Herzer97}.
Thus, understanding and controlling polycrystalline structures is of great 
importance in the production of many engineering materials and has 
motivated numerous experimental and theoretical studies 
of grain growth. 

\begin{figure}[ht] \centering
\includegraphics[width=0.222\textwidth]{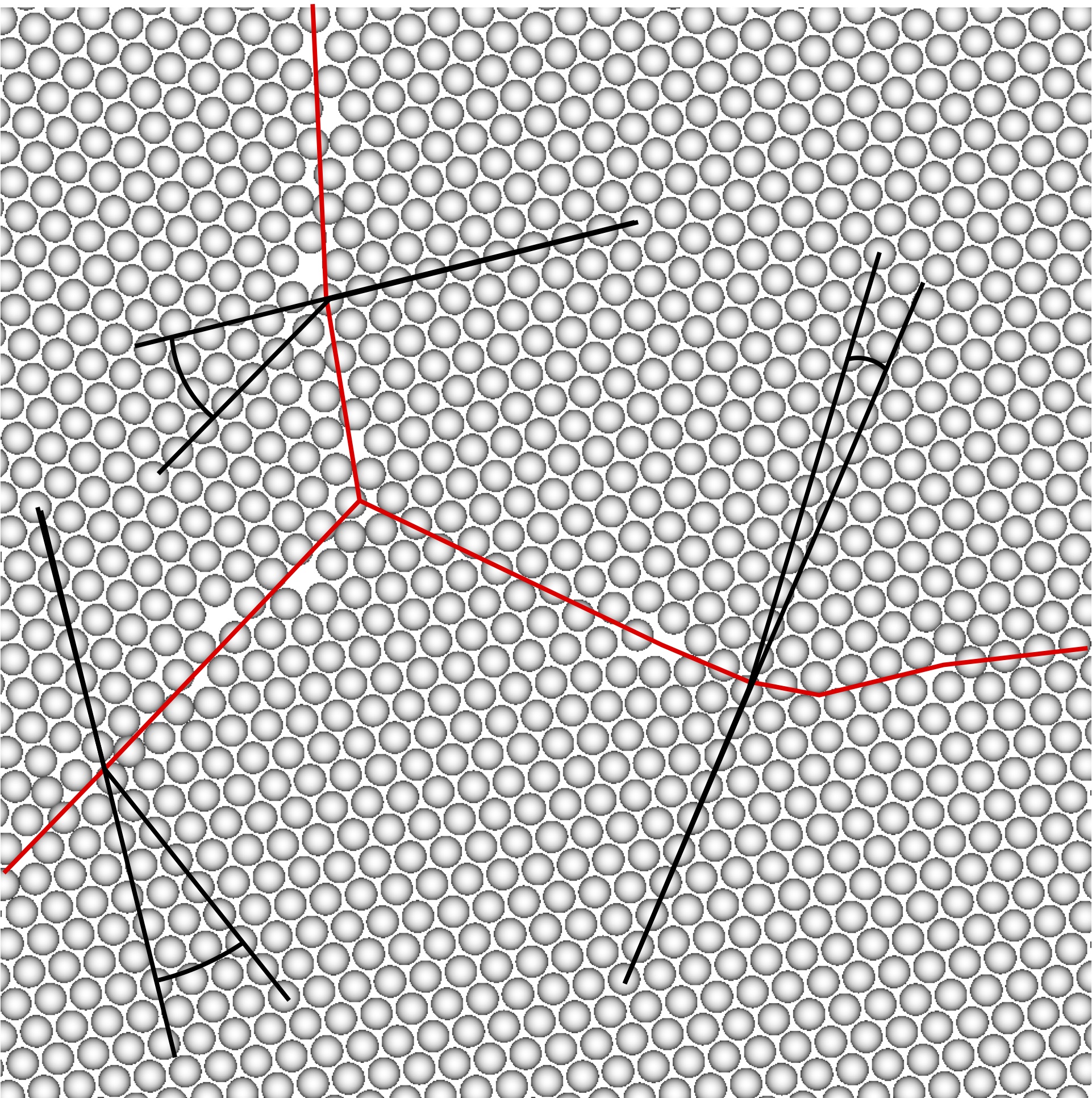} 
\includegraphics[width=0.221\textwidth]{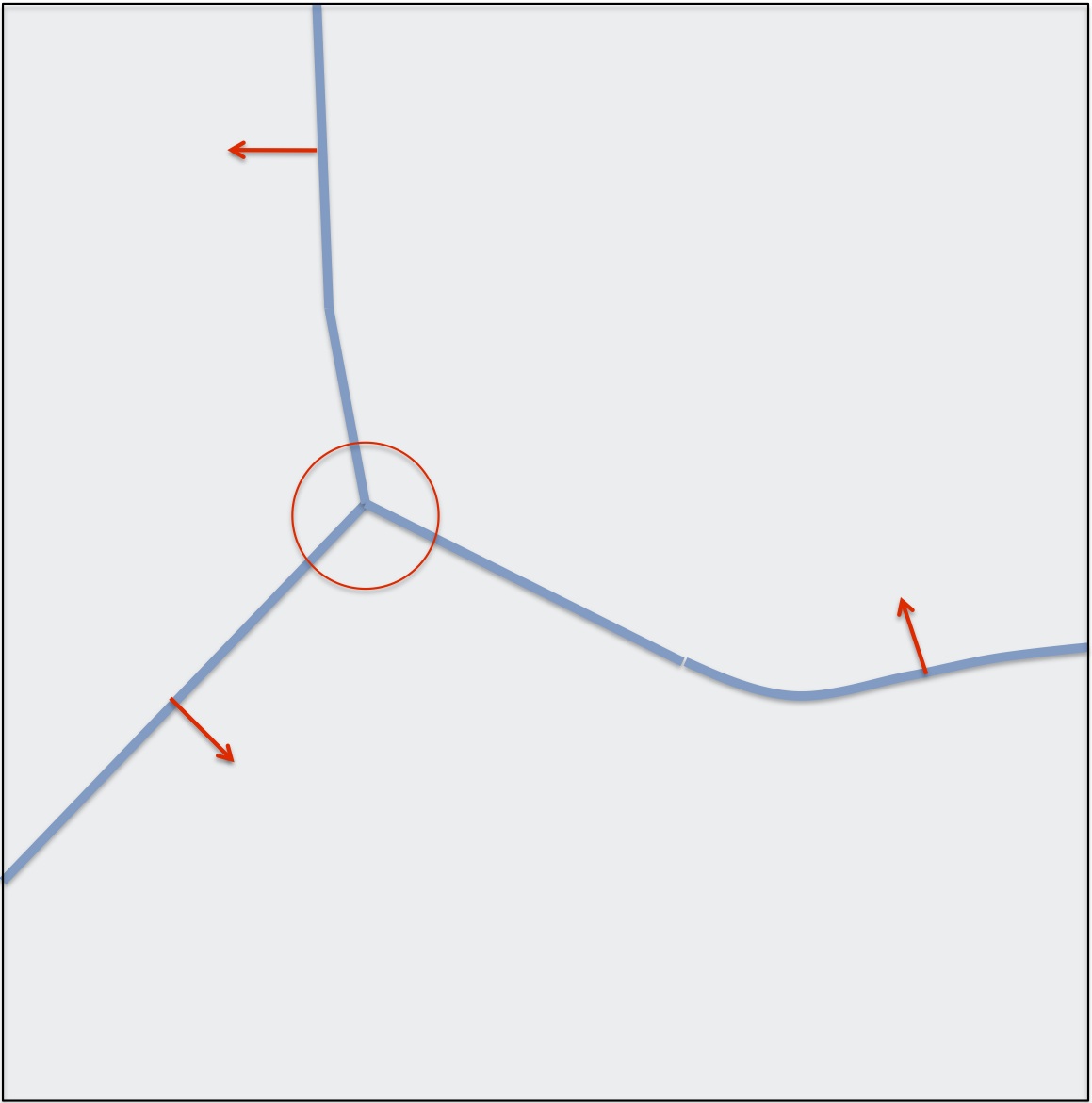} 
\caption{Schematic comparison between atomistic description of polycrystalline
material and coarse grained picture of a smooth grain boundary network. Shown
is a low angle grain boundary with aligned dislocations and two high angle
grain boundaries in an otherwise hexagonal lattice.}
\label{fig1}
\end{figure}

\begin{figure*}[t] \centering
\includegraphics[width=0.95\textwidth]{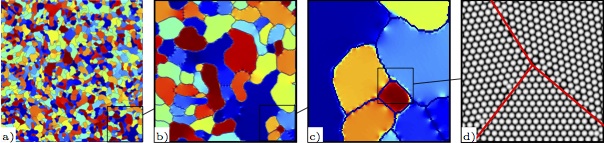} 
\caption{a) - c) Grain structure obtained from postprocessing a PFC simulation
at an intermediate time. The color coding indicates the averaged local lattice
orientation for each of the maxima in the density field. An enlargement by a
factor of four is used for each figure. The grain boundaries are visualized. d)
Visualization of the density field, where each maximum is associated with a
dot. The red lines indicate the grain boundaries.} \label{fig2}
\end{figure*}

Grain growth in thin metallic films is one example where extensive research has 
been conducted. One very interesting experimental finding in such systems is 
that the grain size distributions and topological characteristics appear to be 
independent of many experimental conditions \cite{Barmaketal_PMS_2013}.
More specifically, it has been found that for a large collection of Al and 
Cu thin films a universal grain size distribution emerges that is 
independent of the substrate, annealing temperature, purity, thickness and 
annealing time.  Unfortunately the universal distribution is qualitatively and quantitatively different from the results of extensive computational studies on grain growth \cite{Elseyetal_PRSA_2011,Kinderlehreretal_SIAMJSC_2006}, which are based on the Mullins model \cite{Mullins_JAP_1956}. In this model the problem is reduced to the evolution of a grain boundary network by relating the normal
velocity $v_n$ to the curvature $\kappa$ of the grain boundary, $v_n = \mu
\gamma \kappa$, with mobility $\mu$ and surface tension $\gamma$, and
specifying the Herring condition \cite{Herring_1951} at triple junctions.
Various attempts have been made to extend the Mullins model and to include more
realistic effects, such as interactions of the film with the 
substrate, anisotropy in the grain boundary energy and mobility,  grain boundary
grooving, and solute and triple junction drag.  
However, no single 
cause is able to explain the experimental grain growth 
behavior \cite{Barmaketal_SM_2006,Barmaketal_PMS_2013}.  

In addition to the grain size distribution, the rate of growth of the average
grain size has also been examined in detail. The Mullins model and its
extensions all seem to predict that the average grain size, represented by its
radius $r(t)$ has a power 
law behavior of the form $\sim t^{1/2}$, which follows immediately from the
linear relationship between grain boundary velocity and curvature.
Experimentally a much slower coarsening or even stagnation 
of grain growth in thin films is observed.  This may be due to the fact 
that the Mullins model ignores the crystalline structure of the grains, the 
dissipation due to lattice deformations and the Peierls barriers for dislocation 
motion.  It is difficult to reconcile Mullins type models with 
the atomistic features of grain boundaries, which (for low angles) can 
be seen as an alignment of dislocations where the driving force for 
grain growth is the stress associated 
with dislocation motion. The differences of the description are shown schematically 
in Fig. \ref{fig1}.  

	Atomistic descriptions can incorporate the important 
physical features missing in the Mullins model and have led to some 
important observations.  It has been shown that the   
complex dislocation structure along curved grain boundaries gives rise to
a misorientation-dependent mobility \cite{Winningetal_AM_2001}. Further studies
indicate that grain boundaries undergo thermal roughening associated with an
abrupt mobility change, leading to smooth (fast) and rough (slow) boundaries
\cite{Holmetal_Science_2010}, which can eventually lead to stagnation of the
growth process. The defect structure at triple junctions can lead to a
sufficiently small mobility limiting the rate of grain boundary migration
\cite{Srinivasanetal_AM_1999,Upmanyuetal_AM_2002}. Also tangential motion of
the lattices are possible. For low-angle grain boundaries, normal and tangential
motion are strongly coupled as a result of the geometric constraint that the
lattices of two crystals change continuously across the interface while the
grain boundary moves \cite{Cahnetal_AM_2004}. 
As a consequence of this coupling, grains rotate as they shrink 
which leads to an increase in the grain boundary energy per unit length, 
although the overall energy decreases since the size of the boundary decreases
\cite{Shanetal_Science_2004,Upmanyuetal_AM_2006,Trauttetal_AM_2012}. Each of
these phenomena can be simulated using molecular dynamics (MD), see
\cite{Mishinetal_AM_2010} for a review. However, to study the effect of these
phenomena on scaling laws, grain size distributions or stagnation of growth
requires a method which operates on diffusive time scales. 

	In this work, we employ the phase field crystal (PFC) method 
\cite{Elderetal_PRL_2002}. The model has been 
shown to successfully model grain boundary energies as a function 
of misorientation \cite{Jaatinenetal_PRE_2009} and non-classical 
grain rotation during grain shrinkage and drag of triple 
junctions \cite{Wuetal_AM_2012}.  In addition 
lower coarsening exponents were already observed for hexagonal 
lattices \cite{Boyeretal_PRL_2002,Ohnogietal_PRE_2011,KarmaPRE}.
The aim of this Letter is to use the PFC model on large scales to obtain
statistical data for scaling laws and grain size distributions and to compare
them with simulation results of the Mullins model and experimental data for
thin metallic films. Since the experimental results 
in \cite{Barmaketal_PMS_2013} seem to be universal,
we do not fit the PFC parameters to a specific material but consider an
artificial setting within the simplest PFC model introduced in
\cite{Elderetal_PRL_2002}. In dimensionless form the equation reads $\partial_t
\psi = \Gamma \nabla^2 \frac{\delta {\cal{F}}}{\delta \psi}$, where the order
parameter $\psi$, is related to the time-averaged atomic density, 
$t$ is time, $\Gamma$ is the mobility and free energy is ${\cal{F}} = \int
\psi(- \epsilon + (\nabla^2 + 1)^2 \psi/2 + \psi^4/4) \; dr$, see
\cite{Elderetal_PRE_2007,Wuetal_PRB_2007,Teeffelenetal_PRE_2009,Jaatinenetal_PRE_2009}
for details on the relationship to classical dynamic density functional theory
(DDFT) and material specific parameterization. 

\begin{figure}[ht]
\centering
\includegraphics[width=0.45\textwidth]{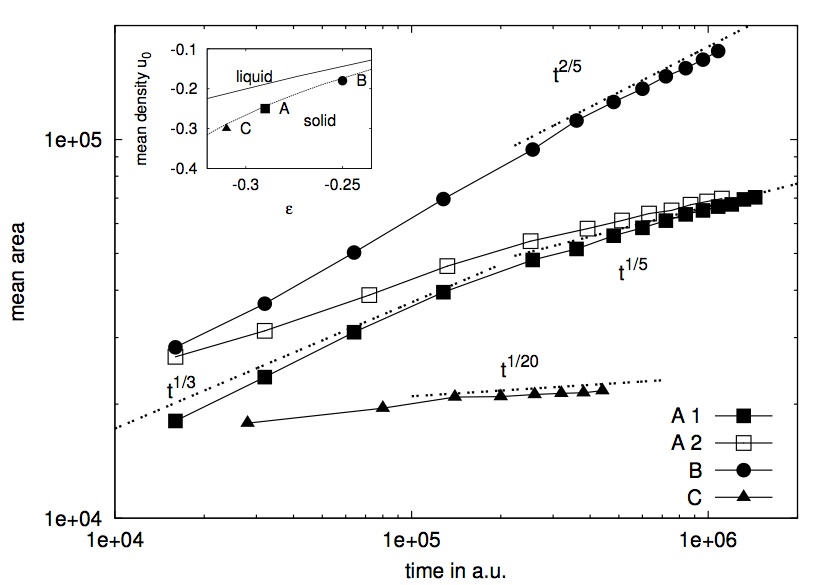} 
\caption{Mean area as a function of time together with the fitted scaling
exponents for various points in the phase diagram depicted in the inset. "A1"
and "A2" have
different initial grain sizes (A1 $<$ A2), the parameters are "A": $(\psi_0,\epsilon) =
(-0.29,-0.25)$; "B": $(\psi_0,\epsilon) = (-0.25,-0.18)$; "C":
$(\psi_0,\epsilon) = (-0.31,-0.30)$.} 
\label{fig3}
\end{figure}

Fig. \ref{fig2} shows a snapshot of a typical simulation.  
All simulations are performed in a periodic domain of square size 
$L = 8,192$
starting from a randomly perturbed constant value of the particle density
$\psi$. 
After an initiation phase in which the white noise is damped rapidly, 
grains nucleate, grow and impinge on one another.
Thereafter the
number of maxima in the particle density $\psi$ remains mainly constant and
coarsening starts. Statistical results are collected after grains have reached
a minimal size of 100 atoms. Fig. \ref{fig3} shows the obtained scaling
results for the average domain area as a power law in times, i.e., $t^{q}$, 
where $q$ is $1/2$ in the Mullins curvature driven model. In our 
simulations it is not clear that this relationship is valid as 
the value of $q$ can be see to change in time and be dependent on 
the parameters of the simulation and initial conditions.
For case "A",  we either obtain an 
initial value of $q = 1/3$, which turns into $q = 1/5$, or a 
constant value of $q = 1/5$, depending on the initial grain size.
The constant scaling exponent is observed for larger initial grains. 
For case "B", corresponding to a softer material, the growth 
exponent increases to a value of $q = 2/5$, 
whereas for case "C", a harder material it decreases to
$q = 1/20$. For
all three cases, the growth exponent is significantly lower than 
the expected value $q = 1/2$ for the Mullins model. Similar low 
coarsening exponents have been found for hexagonal lattices 
in \cite{Boyeretal_PRL_2002,Ohnogietal_PRE_2011}
and in experiments for thin films of CoPt and FePt \cite{Ristauetal_JMR_1999}. 
Extensive computational studies in \cite{Ohnogietal_PRE_2011,KarmaPRE} further 
show a strong dependency of the scaling exponent on additional noise, 
which enhances the coarsening process.  It has also been noted \cite{KarmaPRE} 
that the addition of higher order time derivatives can change 
the growth exponent, which may be appropriate for three dimensional samples. 
In two-dimensional thin films (i.e., films with columnar grain structures), 
however, it is expected that the substrate/film coupling provides an 
effective friction for rotation or translation that eliminates the need 
for such corrections. 
In either case, it is likely that the growth exponents are transient, 
because for very large gain sizes the Peierls-Nabarro barriers are likely 
to inhibit further coarsening.  This effect already occur at early times 
for quenches to lower temperatures, as confirmed for points in the 
phase diagram in the solid region at $(\psi_0, \epsilon) = (-0.31,-0.25)$ and 
$(\psi_0, \epsilon) = (-0.29,-0.18)$  
which show a frozen configuration.  

\begin{figure}[t]
\centering
\includegraphics[width=0.49\textwidth]{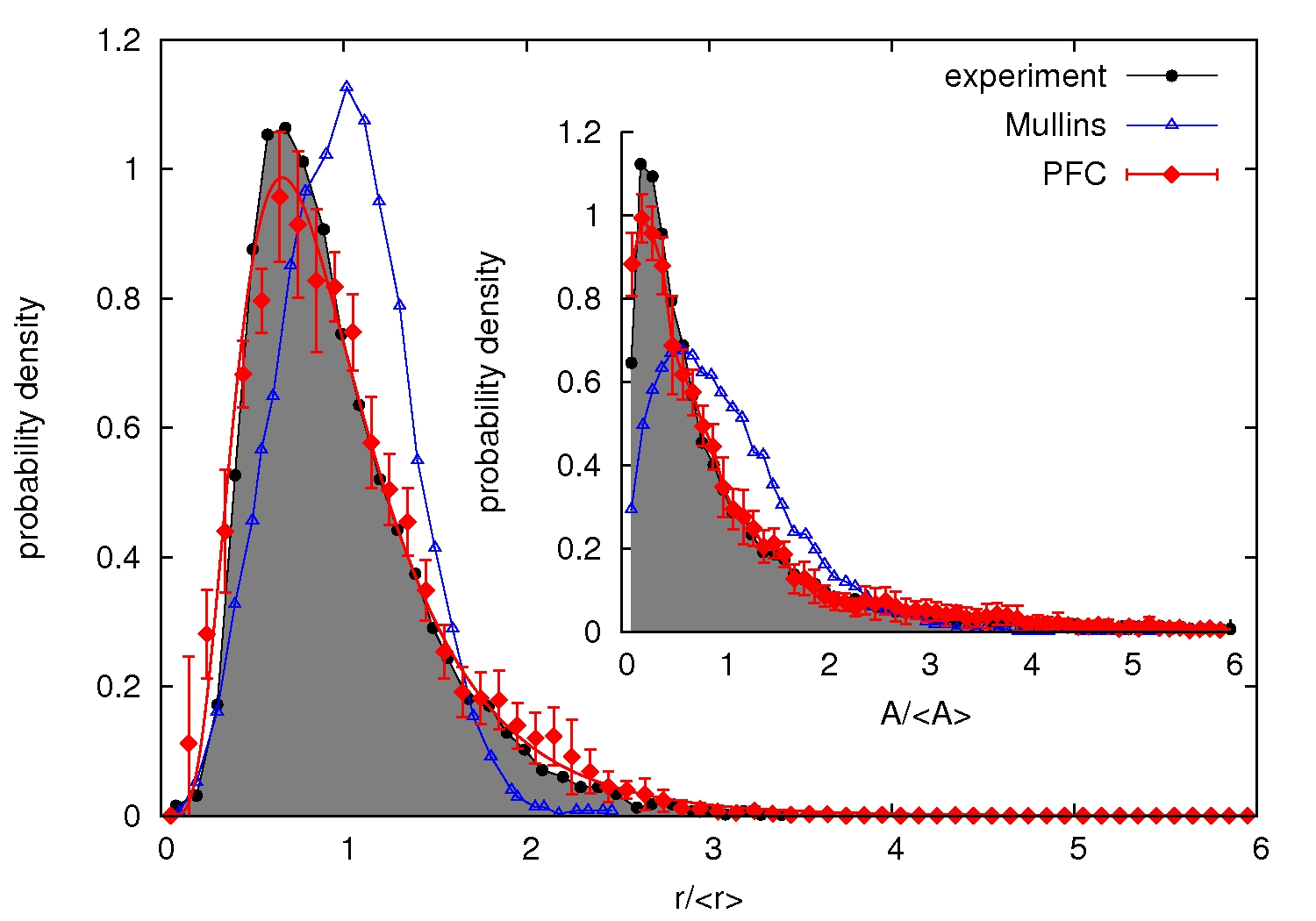} 
\caption{Grain size distribution with reference to radius (area in inlet). Shown is the mean distribution, obtained as the average of the last time steps in the considered cases in Fig. \ref{fig3}. The curve are fitted to a lognormal distribution with parameters $(\mu, \sigma) = (-0.13, 0.53)$. The experimental data and the results of the Mullins model are taken from \cite{Barmaketal_PMS_2013}}. \label{fig4}
\end{figure}

While it is not entirely clear if there is a single, well-established 
dynamical exponent,
the grain size distribution functions appear to be much more robust. Fig.
\ref{fig4} shows the grain size distributions of the PFC simulations for the
considered points in the phase diagram together with the experimental results
of \cite{Barmaketal_PMS_2013} and the results of the Mullins model taken from 
\cite{Elseyetal_PRSA_2011}. 

A considerable discrepancy between the experimental 
results and the Mullins model is already discussed in
\cite{Barmaketal_SM_2006,Barmaketal_PMS_2013}.  They differ in two important
respects. First, the experimental grain structures have a larger 
number of small grains as evidenced by the peak of the experimental reduced
area probability density residing to the left of that for the simulations based
on the Mullins model, a feature that has been 
termed the "ear". Second, the experimental grain structures have "tails" that
extend to significantly larger sizes than those seen in simulations based on
the Mullins model. While only very few grains 
seen in simulations exceed 4 times (and only rarely do they exceed 5 times) the
mean area, the experimental grain structures exhibit maximum grain areas that
are between 8 and 42 times the 
mean, with a sizable fraction of grains whose areas exceed 4 times the mean
grain area ($\sim 3\%$ by number, representing $\sim 18\%$ of the total area).

\begin{figure}[h]
\centering
a) \includegraphics[width=0.46\textwidth]{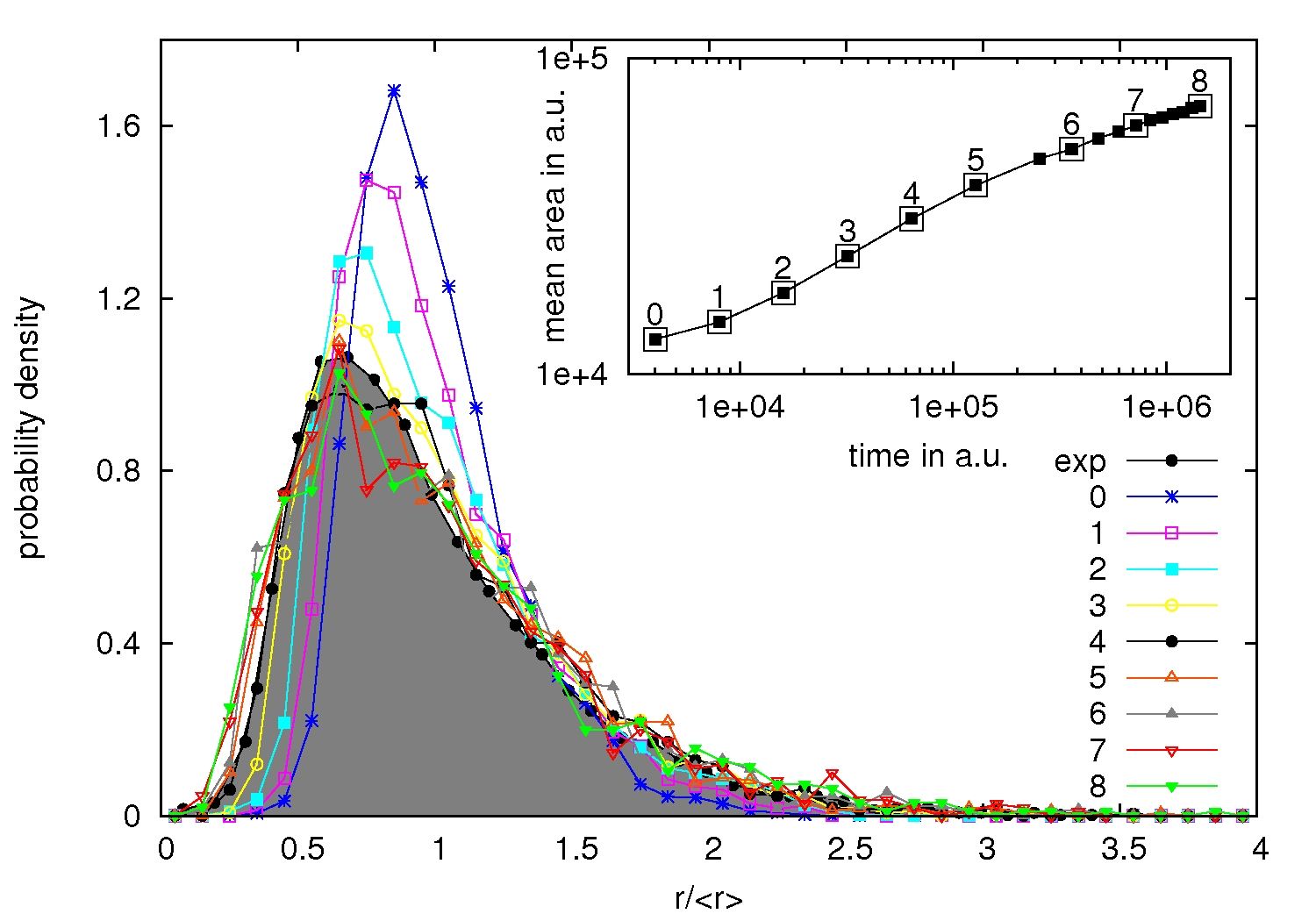} 
b) \includegraphics[width=0.4\textwidth]{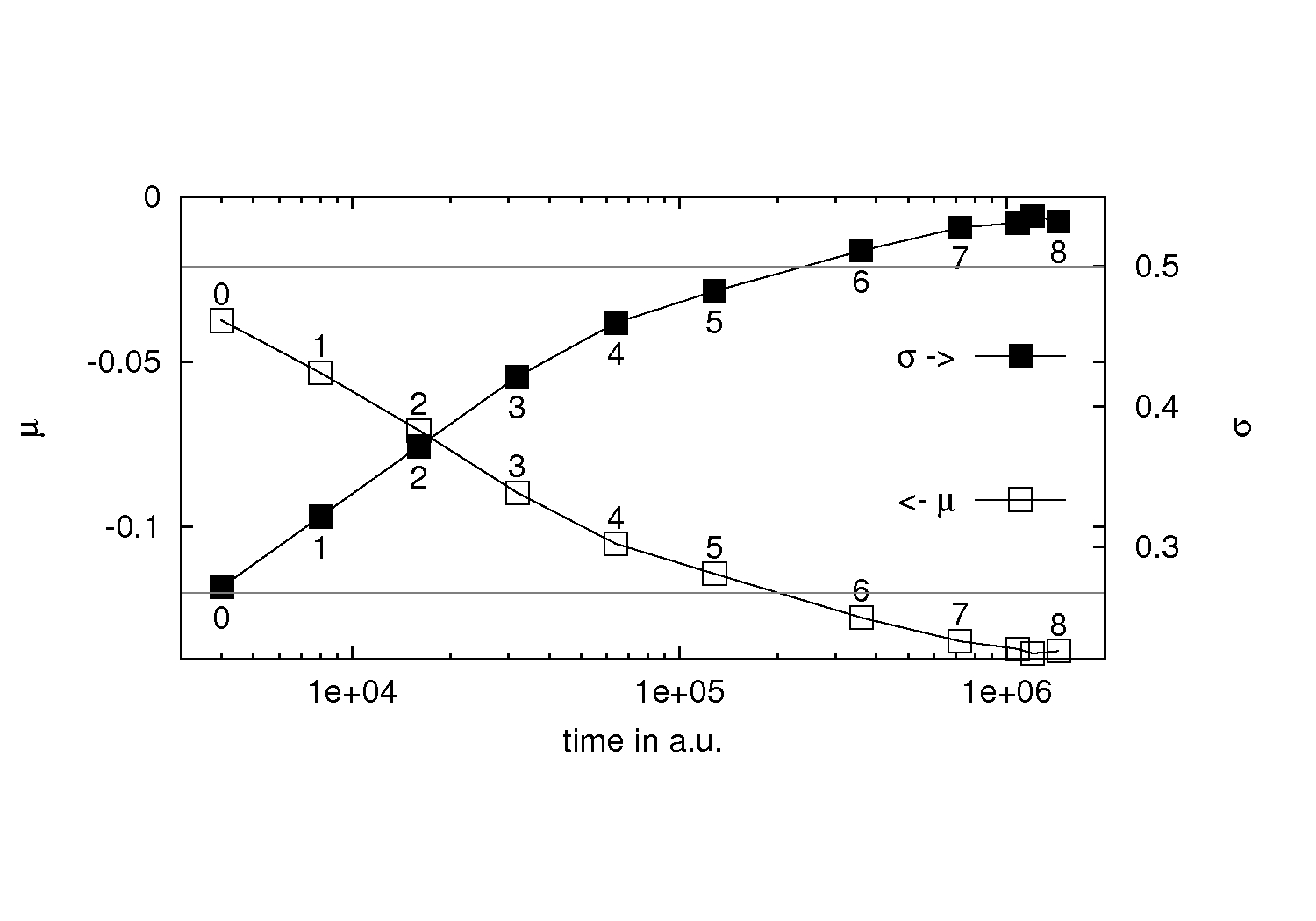} 
\caption{a) Grain size distribution with reference to radius at the labled times in
the inlet, cooresponding to case "A1" in Fig. \ref{fig3}, in comparison with the 
experimental results from \cite{Barmaketal_PMS_2013}. The initially narrow distribution
broadens rapidly and its peak shifts towards smaller
grains. For large times the grain size distribution appears to be self-similar
which is further illustrated in b) showing the time evolution of the
parameters $\sigma$ and $\mu$ of a log-normal distribution fitted to the 
considered snapshots, again in comparison with the experimental results
from \cite{Barmaketal_PMS_2013} shown as the horizontal solid lines.
} 
\label{sfig1}
\end{figure}

Various closed form distributions have been proposed to fit the results of the
Mullins model, e.g. the Louat, Hillert, Rios and Weibull distribution (see
\cite{Elseyetal_PRSA_2011} and the references therein). They all not only 
differ in the "ear" and "tail" region, but they also peak at $r /
\langle r \rangle > 1$, again in disagreement with the experimental results. The
PFC simulations not only recover the qualitative behaviour of the experimental 
results, they almost perfectly fit the distribution, and can be very well described 
by a log-normal distribution. The grain size distribution appears to 
be self-similar which is analysed in detail for case "A1" in 
Fig. \ref{sfig1}. All results are obtained without 
additional noise. However, simulations that included noise (not shown) produced grain distributions
consistent with the zero noise case.  Further analysis indicates that, also in
agreement with the experimental data, small grains are primarily 3 and 4-sided,
whereas large grains have primarily more than 6 sides.

	The importance and prevalence of the formation and 
properties of polycrystalline materials has lead to an enormous 
amount of theoretical and experimental research. Unfortunately 
theoretical progress has been hindered by the lack of computational 
methods that can capture the essential physics on the time and 
lengths that are appropriate for such phenomena. While MD simulations
are currently unable to reach time scales required to observe self-similar
growth regimes, coarse grained descriptions based on the Mullins model
seem to lack the essential atomistic features allowing for bulk dissipation
during grain growth. In this work large scale numerical simulations of the PFC model 
were used to examine the phenomenon of grain growth in two dimensional
systems. The results of these simulations are in remarkable agreement 
with universal aspects of the geometric and topological characteristics 
of the grain structures in thin metallic films. Among other features they 
capture both the "ear" and "tail" characteristics of grain distributions that 
have proven difficult to obtain with previous models and methods. 
Thus the PFC model provides a key resource for future research in 
which realistic grain structures are required.  Although not 
examined in this work, the model also incorporates mechanical 
properties of the system and thus can be used to study, for 
example, the relationship between growth conditions and the 
structural stability of polycrystalline materials. 

\begin{acknowledgments}
RB and AV acknowledge support from the DFG under Grant No. Vo899/7. KE acknowledges support from the NSF under Grant No. DMR-0906676. We acknowledge computing resources at the JSC provided under grant HDR06. Part of the work has been done while AV was guest of HIM at Universit\"at Bonn and KB, KE and AV were guests of IPAM at UCLA. 
\end{acknowledgments}
    

\bibliography{lit2}

\end{document}